\documentstyle[12pt]{article}
\title{Singular solutions of Yang-Mills equations \\
 and bag model.}

\author{
F. A. Lunev 
 \thanks{e-mail address: lunev@hep.phys.msu.su }
\\ {\em Physical Department, Moscow
 State University,}\\ {\em Moscow, 119899, Russia}
\and O. V. Pavlovsky
 \thanks{e-mail address: pavlovsky@hep.phys.msu.su }
\\ {\em Physical Department, Moscow
 State University,}\\ {\em Moscow, 119899, Russia}
 }

\date{ \ \ \  }

\begin{document}

\maketitle

\begin{abstract}
A model of quark confinement based on a singular solution of 
classical YM equation is proposed. Within the framework of this 
model we have calculated hadron masses that correspond to ground 
state configurations of quarks. Our results are in agreement with 
the experiment data with accuracy 3-7 percents for all hadronic 
masses except those of light pseudoscalar mesons.
\end{abstract}

\vspace{1.5cm}

\section{Introduction.}
 Bag model is one of the first attempts to understand physics 
of hadrons in terms of quarks. In its simplest form it was 
formulated in the end of 60's \cite{Bogol} and became well-known 
after the works of MIT group \cite{MIT} in the middle of 70's. 
So far many modifications of bag model were proposed \cite{Moddif}. 
It appeared that bag model gave rather satisfactory description of 
masses, magnetic moments and many other properties of hadrons 
\cite{Propot}. However, in spite of these successes, bag model never 
was a very popular one. The origin is clear. In contrast to Standard 
model pretending for description of hadron physics 
by proceeding
from some fundamental principles, bag model is based on a pure 
phenomenological assumption that quarks are confined inside a 
certain sphere. Up to now it is absolutely unclear how to derive 
such an assumption from QCD and just so nowadays
bag model is usually considered as a rather crude phenomenological 
model, and nothing more.

In the present paper we will try to show that , nevertheless, 
there exists a way to derive a kind of bag model from QCD.

Our basic assumption is that quark in zero approximation moves in 
a certain effective YM potential, that is a solution to classical YM 
equations with singularity on the sphere. Such solutions were 
discovered in 70's in papers \cite{Swank}. More recently 
singular solution of YM equations was found as an analogue to 
Schwartzchild solution in general relativity \cite{Lunev93}. In the 
latter paper a remarkable analogy between YM theory and general 
relativity established in the works 
\cite{Lunev92} was used. (See \cite{Lunev95} 
for further references). Later analogous solutions for 
Yang-Mills-Higgs equations were found \cite{Sing}.  Solutions 
with singularity on the torus and cylinder were investigated in  
\cite{Obukhov}.

Our model obviously can be derived from QCD by quantization in 
neighborhood of such a singular classical solution as zero 
approximation. Further corrections, in principal, can be also 
obtained in a systematic way. But in present paper we restrict 
ourselves to investigation of zero approximation. Namely, we will 
investigate the motion of quark in such field and evaluate mass 
spectrum of ground hadron states.  We will show that our model gives 
quite satisfactory description of all ground hadron state except 
those of light pseudoscolar mesons. The discrepancy between our 
results and experiment data for masses of light pseudoscolar mesons 
is not a surprise. Indeed, it is known that many features of 
light mesons physics are related to spontaneous breaking of chiral 
symmetry. But the latter phenomenon is not taken into consideration 
within framework of our bag model at all. So one cannot hope to 
describe light meson mass spectrum within framework of the present 
simple version of our model. This problem needs further 
investigations.

The paper is organized as follows. In section 2 we describe   
spherically symmetric solutions of YM equations with singularity 
on the sphere. In section 3 we investigate motion of Dirac particle 
in such field. In section 4 we calculate the mass spectrum of ground 
hadron states by quantization in the neighborhood of this classical 
solution. In the last section we discuss results obtained.

\section{Classical solutions of YM equations with  
singularity on the sphere. }

Let us consider four dimensional SU(2) YM theory.  
Substituting the well-known Wu-Yang ansatz \cite{wy}
 $$
   A^a_0=0,\,
A_i^a=\varepsilon_{aij} {{x^j}\over {r^2}}(1-H)\eqno(1)
 $$
in YM equations, one get
 $$ 
r^2H^{''} =H(H^2-1)\eqno(2) 
 $$

It can be proved \cite{Swank}
that there are only two types of solutions of equation 
(2) that are singular on some sphere and regular at $r=0$
(see Fig.1).

\begin{figure}[tbh]
\begin{center}
\unitlength=1mm

\begin{picture}(80,110)

\put(0,0){\line(1,0){80}}
\put(0,110){\line(1,0){80}}
\put(0,0){\line(0,1){110}}
\put(80,0){\line(0,1){110}}

\put(0,10){
   \begin{picture}(80,100)
   \put(10,5){\line(1,0){69}}
   \put(10,0){\line(0,1){100}}
\put(10.7000,10.0015){\line(1,0){0.1}} \put(10.7000,0.0173){\line(1,0){0.1}}
\put(11.4000,10.0061){\line(1,0){0.1}} \put(11.4000,0.0628){\line(1,0){0.1}}
\put(12.1000,10.0137){\line(1,0){0.1}} \put(12.1000,0.1362){\line(1,0){0.1}}
\put(12.8000,10.0243){\line(1,0){0.1}} \put(12.8000,0.2594){\line(1,0){0.1}}
\put(13.5000,10.0381){\line(1,0){0.1}} \put(13.5000,0.3891){\line(1,0){0.1}}
\put(14.2000,10.0548){\line(1,0){0.1}} \put(14.2000,0.5427){\line(1,0){0.1}}
\put(14.9000,10.0747){\line(1,0){0.1}} \put(14.9000,0.7183){\line(1,0){0.1}}
\put(15.6000,10.0978){\line(1,0){0.1}} \put(15.6000,0.9212){\line(1,0){0.1}}
\put(16.3000,10.1239){\line(1,0){0.1}} \put(16.3000,1.1401){\line(1,0){0.1}}
\put(17.0000,10.1533){\line(1,0){0.1}} \put(17.0000,1.3767){\line(1,0){0.1}}
\put(17.7000,10.1858){\line(1,0){0.1}} \put(17.7000,1.6317){\line(1,0){0.1}}
\put(18.4000,10.2216){\line(1,0){0.1}} \put(18.4000,1.9012){\line(1,0){0.1}}
\put(19.1000,10.2607){\line(1,0){0.1}} \put(19.1000,2.1826){\line(1,0){0.1}}
\put(19.8000,10.3031){\line(1,0){0.1}} \put(19.8000,2.4744){\line(1,0){0.1}}
\put(20.5000,10.3488){\line(1,0){0.1}} \put(20.5000,2.7750){\line(1,0){0.1}}
\put(21.2000,10.3981){\line(1,0){0.1}} \put(21.2000,3.0828){\line(1,0){0.1}}
\put(21.9000,10.4508){\line(1,0){0.1}} \put(21.9000,3.3965){\line(1,0){0.1}}
\put(22.6000,10.5070){\line(1,0){0.1}} \put(22.6000,3.7146){\line(1,0){0.1}}
\put(23.3000,10.5669){\line(1,0){0.1}} \put(23.3000,4.0359){\line(1,0){0.1}}
\put(24.0000,10.6305){\line(1,0){0.1}} \put(24.0000,4.3594){\line(1,0){0.1}}
\put(24.7000,10.6979){\line(1,0){0.1}} \put(24.7000,4.6842){\line(1,0){0.1}}
\put(25.4000,10.7691){\line(1,0){0.1}} \put(25.4000,5.0094){\line(1,0){0.1}}
\put(26.1000,10.8442){\line(1,0){0.1}} \put(26.1000,5.3342){\line(1,0){0.1}}
\put(26.8000,10.9234){\line(1,0){0.1}} \put(26.8000,5.6581){\line(1,0){0.1}}
\put(27.5000,11.0068){\line(1,0){0.1}} \put(27.5000,5.9807){\line(1,0){0.1}}
\put(28.2000,11.0944){\line(1,0){0.1}} \put(28.2000,6.3016){\line(1,0){0.1}}
\put(28.9000,11.1864){\line(1,0){0.1}} \put(28.9000,6.6205){\line(1,0){0.1}}
\put(29.6000,11.2829){\line(1,0){0.1}} \put(29.6000,6.9374){\line(1,0){0.1}}
\put(30.3000,11.3841){\line(1,0){0.1}} \put(30.3000,7.2520){\line(1,0){0.1}}
\put(31.0000,11.4900){\line(1,0){0.1}} \put(31.0000,7.5645){\line(1,0){0.1}}
\put(31.7000,11.6009){\line(1,0){0.1}} \put(31.7000,7.8748){\line(1,0){0.1}}
\put(32.4000,11.7169){\line(1,0){0.1}} \put(32.4000,8.1830){\line(1,0){0.1}}
\put(33.1000,11.8381){\line(1,0){0.1}} \put(33.1000,8.4894){\line(1,0){0.1}}
\put(33.8000,11.9648){\line(1,0){0.1}} \put(33.8000,8.7941){\line(1,0){0.1}}
\put(34.5000,12.0972){\line(1,0){0.1}} \put(34.5000,9.0975){\line(1,0){0.1}}
\put(35.2000,12.2355){\line(1,0){0.1}} \put(35.2000,9.3997){\line(1,0){0.1}}
\put(35.9000,12.3799){\line(1,0){0.1}} \put(35.9000,9.7013){\line(1,0){0.1}}
\put(36.6000,12.5306){\line(1,0){0.1}} \put(36.6000,10.0024){\line(1,0){0.1}}
\put(37.3000,12.6879){\line(1,0){0.1}} \put(37.3000,10.3037){\line(1,0){0.1}}
\put(38.0000,12.8522){\line(1,0){0.1}} \put(38.0000,10.6054){\line(1,0){0.1}}
\put(38.7000,13.0236){\line(1,0){0.1}} \put(38.7000,10.9082){\line(1,0){0.1}}
\put(39.4000,13.2026){\line(1,0){0.1}} \put(39.4000,11.2124){\line(1,0){0.1}}
\put(40.1000,13.3894){\line(1,0){0.1}} \put(40.1000,11.5186){\line(1,0){0.1}}
\put(40.8000,13.5845){\line(1,0){0.1}} \put(40.8000,11.8274){\line(1,0){0.1}}
\put(41.5000,13.7882){\line(1,0){0.1}} \put(41.5000,12.1393){\line(1,0){0.1}}
\put(42.2000,14.0010){\line(1,0){0.1}} \put(42.2000,12.4549){\line(1,0){0.1}}
\put(42.9000,14.2233){\line(1,0){0.1}} \put(42.9000,12.7749){\line(1,0){0.1}}
\put(43.6000,14.4556){\line(1,0){0.1}} \put(43.6000,13.1001){\line(1,0){0.1}}
\put(44.3000,14.6986){\line(1,0){0.1}} \put(44.3000,13.4315){\line(1,0){0.1}}
\put(45.0000,14.9527){\line(1,0){0.1}} \put(45.0000,13.7694){\line(1,0){0.1}}
\put(45.7000,15.2185){\line(1,0){0.1}} \put(45.7000,14.1148){\line(1,0){0.1}}
\put(46.4000,15.4969){\line(1,0){0.1}} \put(46.4000,14.4684){\line(1,0){0.1}}
\put(47.1000,15.7885){\line(1,0){0.1}} \put(47.1000,14.8309){\line(1,0){0.1}}
\put(47.8000,16.0942){\line(1,0){0.1}} \put(47.8000,15.2032){\line(1,0){0.1}}
\put(48.5000,16.4148){\line(1,0){0.1}} \put(48.5000,15.5881){\line(1,0){0.1}}
\put(49.2000,16.7512){\line(1,0){0.1}} \put(49.2000,15.9849){\line(1,0){0.1}}
\put(49.9000,17.1047){\line(1,0){0.1}} \put(49.9000,16.3943){\line(1,0){0.1}}
\put(50.6000,17.4762){\line(1,0){0.1}} \put(50.6000,16.8202){\line(1,0){0.1}}
\put(51.3000,17.8672){\line(1,0){0.1}} \put(51.3000,17.2613){\line(1,0){0.1}}
\put(52.0000,18.2790){\line(1,0){0.1}} \put(52.0000,17.7211){\line(1,0){0.1}}
\put(52.7000,18.7130){\line(1,0){0.1}} \put(52.7000,18.1993){\line(1,0){0.1}}
\put(53.4000,19.1712){\line(1,0){0.1}} \put(53.4000,18.7006){\line(1,0){0.1}}
\put(54.1000,19.6553){\line(1,0){0.1}} \put(54.1000,19.2245){\line(1,0){0.1}}
\put(54.8000,20.1674){\line(1,0){0.1}} \put(54.8000,19.7738){\line(1,0){0.1}}
\put(55.5000,20.7099){\line(1,0){0.1}} \put(55.5000,20.3515){\line(1,0){0.1}}
\put(56.2000,21.2855){\line(1,0){0.1}} \put(56.2000,20.9603){\line(1,0){0.1}}
\put(56.9000,21.8970){\line(1,0){0.1}} \put(56.9000,21.6034){\line(1,0){0.1}}
\put(57.6000,22.5478){\line(1,0){0.1}} \put(57.6000,22.2839){\line(1,0){0.1}}
\put(58.3000,23.2415){\line(1,0){0.1}} \put(58.3000,23.0048){\line(1,0){0.1}}
\put(59.0000,23.9825){\line(1,0){0.1}} \put(59.0000,23.7728){\line(1,0){0.1}}
\put(59.7000,24.7755){\line(1,0){0.1}} \put(59.7000,24.5904){\line(1,0){0.1}}
\put(60.4000,25.6260){\line(1,0){0.1}} \put(60.4000,25.4644){\line(1,0){0.1}}
\put(61.1000,26.5403){\line(1,0){0.1}} \put(61.1000,26.4015){\line(1,0){0.1}}
\put(61.8000,27.5256){\line(1,0){0.1}} \put(61.8000,27.4088){\line(1,0){0.1}}
\put(62.5000,28.5905){\line(1,0){0.1}} \put(62.5000,28.4942){\line(1,0){0.1}}
\put(63.2000,29.7447){\line(1,0){0.1}} \put(63.2000,29.6685){\line(1,0){0.1}}
\put(63.9000,30.9996){\line(1,0){0.1}} \put(63.9000,30.9439){\line(1,0){0.1}}
\put(64.6000,32.3690){\line(1,0){0.1}} \put(64.6000,32.3326){\line(1,0){0.1}}
\put(65.3000,33.8690){\line(1,0){0.1}} \put(65.3000,33.8524){\line(1,0){0.1}}
\put(66.0000,35.5188){\line(1,0){0.1}} \put(66.0000,35.5232){\line(1,0){0.1}}
\put(66.7000,37.3420){\line(1,0){0.1}} \put(66.7000,37.3676){\line(1,0){0.1}}
\put(67.4000,39.3670){\line(1,0){0.1}} \put(67.4000,39.4152){\line(1,0){0.1}}
\put(68.1000,41.6291){\line(1,0){0.1}} \put(68.1000,41.7013){\line(1,0){0.1}}
\put(68.8000,44.1722){\line(1,0){0.1}} \put(68.8000,44.2719){\line(1,0){0.1}}
\put(69.5000,47.0517){\line(1,0){0.1}} \put(69.5000,47.1821){\line(1,0){0.1}}
\put(70.2000,50.3389){\line(1,0){0.1}} \put(70.2000,50.5051){\line(1,0){0.1}}
\put(70.9000,54.1266){\line(1,0){0.1}} \put(70.9000,54.3349){\line(1,0){0.1}}
\put(71.6000,58.5379){\line(1,0){0.1}} \put(71.6000,58.7982){\line(1,0){0.1}}
\put(72.3000,63.7402){\line(1,0){0.1}} \put(72.3000,64.0658){\line(1,0){0.1}}
\put(73.0000,69.9667){\line(1,0){0.1}} \put(73.0000,70.3762){\line(1,0){0.1}}
\put(73.7000,77.5521){\line(1,0){0.1}} \put(73.7000,78.0734){\line(1,0){0.1}}
\put(74.4000,86.9939){\line(1,0){0.1}} \put(74.4000,87.6712){\line(1,0){0.1}}
\put(75.1000,99.0640){\line(1,0){0.1}} \put(75.1000,99.9719){\line(1,0){0.1}}

\multiput(10,5)(7,0){10}{\line(0,1){1}}
\multiput(10,0)(0,5){20}{\line(1,0){1}}

\put(14.5,1){0.1}
\put(21.5,1){0.2}
\put(28.5,1){0.3}
\put(35.5,1){0.4}
\put(42.5,1){0.5}
\put(49.5,1){0.6}
\put(56.5,1){0.7}
\put(63.5,1){0.8}
\put(70.5,1){0.9}

   \end{picture}}

\put(7,8.5){-1}
\put(8,13.5){0}
\put(8,18.5){1}
\put(8,23.5){2}
\put(8,28.5){3}
\put(8,33.5){4}
\put(8,38.5){5}
\put(8,43.5){6}
\put(8,48.5){7}
\put(8,53.5){8}
\put(8,58.5){9}
\put(7,63.5){10}
\put(7,68.5){11}
\put(7,73.5){12}
\put(7,78.5){13}
\put(7,83.5){14}
\put(7,88.5){15}
\put(7,93.5){16}
\put(2,103.5){\bf H(r)}

\put(25,23){$H_1$}
\put(45,20){$H_2$}

\end{picture}
\end{center}
\caption{The singular solutions of Wu-Yang equation (2),$\, R=1$.} 

\end{figure}
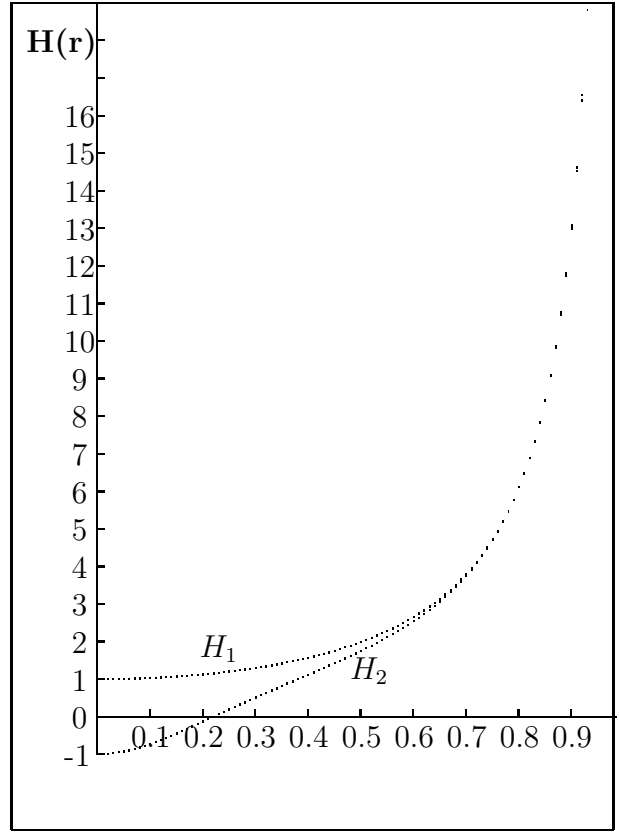

The asymptotics of the first one are: 
  $$
  H_1(r) \sim {\sqrt{2} \over 
 {1-r/R}}\, , \, r\to R-0\, ;\qquad H_1(r) \sim 1+c_1 \Bigl({r \over 
 R}\Bigr)^2 \, , \, r\to 0 \eqno(3)
  $$
  and of the second one are: 
  $$ 
 H_2(r) 
 \sim {\sqrt{2} \over {1-r/
 R}}\, , \, r\to R-0\, ;\qquad H_2(r) \sim 
 -1+c_2 \Bigl({r \over R}\Bigr)^2 \, , \, r\to 0 \eqno(4)
 $$ 
 where $R>0$ is 
 an arbitrary constant, $c_1\simeq 3.038 $ and $c_2\simeq 9.448 $.

It appears that satisfactory agreement between experimental data and 
results obtained from our model can be achieved only for solution 
with asymptotics (4). So in what follows we will suppose that 
functions $H(r)$ is defined by eqs.(2),(4). Starting from these 
singular solutions of $SU(2)$ YM equations, one can construct two 
different sets of singular solutions of $SU(3)$ YM equations that 
correspond to two non-equivalent embeddings of the algebra $SU(2)$ 
in the algebra $SU(3)$. Namely, up to unitary equivalence,
the first set of such solutions can be defined as 
$$ 
 A={1\over 2}A^1\lambda^1+{1 \over 2}A^2\lambda^2+ {1 \over 
2}A^3\lambda^3 \eqno(5) 
$$ 
and the second one as
$$
 A={1 \over 2}A^1\lambda^2+{1 
\over 2}A^2\lambda^5+ {1 \over 2}A^3\lambda^7 \eqno(6) 
$$
 where 
$\lambda^\alpha,\, \alpha=1,2\cdots 8,$ are Gell-Mann matrices and 
$A^a,\, a=1,2,3$ are defined by formula (1).

\section{Quark in Yang-Mills field with singularity on the sphere.}

Let us consider solutions of Dirac equation
  $$
  ( i\gamma^0 {\partial \over {\partial t}} +i 
  {\gamma}^j {\nabla}_j ) \Psi =m\Psi\eqno(7) 
  $$
   for the quark in potential 
  (5) or (6). In the case (5) quark cannot be confined inside the 
  sphere $r=R$ because the third component of quark field satisfies 
  free Dirac equation.  So in  what follows we restrict ourselves to 
  consideration of the case (6).

  By virtue of spherical symmetry of the potential (6) Dirac 
Hamiltonian commutes with  operators of angular momentum
$$
J^a={i \over 4}\varepsilon^{abc} [\gamma^b,\gamma^c]+I^a+l^a
$$
where $(I^a)\equiv (\lambda^2,\lambda^5,\lambda^7)$ are color 
isospin operators and $l^a$ are orbital angular momentum ones,
 $$
 l_i=-i \varepsilon_{ijk} x_j \partial_k
 $$

The solutions with definite total angular momentum and definite 
energy $E$ can be represented as
 $$
 \begin{array}{l}
 \Psi = {1 \over r}
\left( \begin{array}{c}
   B^{+}_1(r)\nabla^{\Omega}_j \Omega^{J+1/2}_{JM}+
   D^{+}_1(r)n_j \Omega^{J+1/2}_{JM}+
   C^{-}_1(r)l_j \Omega^{J-1/2}_{JM}\\
   i\sqrt{{E-m} \over {E+m}}
   (\, B^{-}_1(r)\nabla^{\Omega}_j\Omega^{J-1/2}_{JM}+
       D^{-}_1(r)n_j \Omega^{J-1/2}_{JM}+ 
       C^{+}_1(r)l_j \Omega^{J+1/2}_{JM} \,)
      \end{array} \right)e^{-iEt}  \\
 + {1 \over r}
\left( \begin{array}{c}
   iB^{-}_2(r)\nabla^{\Omega}_j\Omega^{J-1/2}_{JM}
 + iD^{-}_2(r)n_j\Omega^{J-1/2}_{JM}
 + iC^{+}_2(r)l_j\Omega^{J+1/2}_{JM}\\ 
  \sqrt{{E-m} \over {E+m}}
    (\, B^{+}_2(r)\nabla^{\Omega}_j \Omega^{J+1/2}_{JM}+ 
        D^{+}_2(r)n_j \Omega^{J+1/2}_{JM}+ 
        C^{-}_2(r)l_j \Omega^{J-1/2}_{JM} \,)
      \end{array} \right)e^{-iEt}  \\
 \end{array}\eqno(8)
 $$
 where $\Omega^{J \pm 1/2}_{JM}$ are a spherical spinors 
 $$
 \Omega^{J+1/2}_{JM}={{-\sqrt{{{J-M+1} \over {2(J+1)}
                                 }} Y_{J+1/2 \, M-1/2}
                           }
                           \choose
                           {\sqrt{{{J+M+1} \over {2(J+1)}
                                 }} Y_{J+1/2 \, M+1/2}
                           }}
 \qquad
 \Omega^{J-1/2}_{JM}={{\sqrt{{{J+M} \over {2(J+1)}
                                 }} Y_{J-1/2 \, M-1/2}
                           }
                           \choose
                           {\sqrt{{{J-M} \over {2(J+1)}
                                 }} Y_{J-1/2 \, M+1/2}
                           }}
 $$
  and
 $$
 \nabla^{\Omega}_i=
 r{\partial \over {\partial x_i}}-x_i
 {\partial \over {\partial r}} 
  \qquad n_i={x_i \over r} 
 $$

One notes, that $J=1/2,3/2,\cdots$. Substituting (8) in (7), one 
gets:
 $$
  \begin{array}{l}

  \begin{array}{l}
 {B^+_{1}}^{'}={\displaystyle {1 \over {J+1/2}}}
 \Bigl( -(J-1/2)(J+3/2) {\displaystyle {1 \over r}}B^+_{1} +   
  {\displaystyle {H \over r}}D^+_{1} 
   -(J-1/2){\displaystyle {1 \over r}}C^-_{1}\\
 \qquad \qquad  \qquad \qquad \qquad \qquad 
 -{\cal E}(C^+_{1}-(J-1/2)B^-_{1})\Bigr) \qquad (\mbox{all} 
\quad J) \end{array} \\ \\

  \begin{array}{l}
 {B^-_{1}}^{'}={\displaystyle {1 \over {J+1/2}}}
 \Bigl( (J-1/2)(J+3/2) {\displaystyle {1 \over r}}B^-_{1} -   
  {\displaystyle {H \over r}}D^-_{1} 
   -(J+3/2){\displaystyle {1 \over r}}C^+_{1}\\
 \qquad \qquad  \qquad \qquad \qquad \qquad
  -{\cal E}(C^-_{1}+(J-1/2)B^+_{1})\Bigr) \qquad (J>1/2) 
  \end{array} \\ \\

  \begin{array}{l}
 {D^+_{1}}^{'}=-(J+1/2){\displaystyle {1 \over r}}D^+_{1} + 
   (J+3/2){\displaystyle {H \over r}}B^+_{1} 
   -(J-1/2){\displaystyle {H \over r}}C^-_{1}\\
 \qquad \qquad  \qquad \qquad \qquad \qquad 
   +{\cal E}D^-_{1}\qquad \qquad \qquad \qquad 
\qquad (\mbox{all}\quad J) \end{array} \\ \\

  \begin{array}{l}
 {D^-_{1}}^{'}=(J+1/2){\displaystyle {1 \over r}}D^-_{1} + 
   (J+3/2){\displaystyle {H \over r}}C^+_{1} 
   -(J-1/2){\displaystyle {H \over r}}B^-_{1} \\
 \qquad \qquad  \qquad \qquad \qquad \qquad
   -{\cal E}D^+_{1}\qquad \qquad \qquad \qquad \qquad 
  (\mbox{all}\quad J) \end{array} \\ \\

  \begin{array}{l}
 {C^-_{1}}^{'}={\displaystyle {1 \over {J+1/2}}}
 \Bigl( (J-1/2)(J+3/2) {\displaystyle {1 \over r}}C^-_{1} -   
  {\displaystyle {H \over r}}D^+_{1} 
   -(J+3/2){\displaystyle {1 \over r}}B^+_{1}\\
 \qquad \qquad  \qquad \qquad \qquad \qquad
        +{\cal E}(B^-_{1}+(J+3/2)C^+_{1})\Bigr) \qquad (J>1/2)  
  \end{array} \\ \\

  \begin{array}{l}
 {C^+_{1}}^{'}={\displaystyle {1 \over {J+1/2}}}
 \Bigl( -(J-1/2)(J+3/2) {\displaystyle {1 \over r}}C^+_{1} +   
  {\displaystyle {H \over r}}D^-_{1} 
   -(J-1/2){\displaystyle {1 \over r}}B^-_{1}\\
 \qquad \qquad  \qquad \qquad \qquad \qquad 
  +{\cal E}(B^+_{1}-(J-1/2)C^-_{1})\Bigr) \qquad (\mbox{all}\quad J)
 \end{array}
 \end{array}
 \eqno(9) 
$$
 where ${\cal 
E}=\sqrt{E^2-m^2}$, and {\it exactly the same} system of equations 
for functions $B^{\pm}_2$,  $C^{\pm}_2$, $D^{\pm}_2$.  So initial 
Dirac equations are separated into two identical sets of equations. 
This leads to degeneration of all energy levels.
 
 This is not surprises. Indeed, Dirac equations in external 
 chromomagnetic field with property (${\vec A}(-x)=-{\vec A}(x)$) 
 are invariant under parity transformations .  
It is well-known \cite{coo} that this symmetry implies 
$N=1$ sypersymmetry of Dirac equations that, in turn, implies the 
doubling of energy levels. Thus all energy levels appear to be 
degenerate in parity. As a consequence, all hadrons must be 
degenerate in parity.  This is a puzzle of our model. In what 
follows, we postulate that quarks may be in states with only one 
parity.  

  States with definite parity correspond to solutions of Dirac 
eq.(7) for which either 
$
B^{\pm}_1=C^{\pm}_1=D^{\pm}_1=0
$
or
$
B^{\pm}_2=C^{\pm}_2=D^{\pm}_2=0.
$
For definiteness, we choose the second case. 

If $J=1/2$ then only four functions $B^+_1(r)$, 
$D^+_1(r)$, $C^+_1(r)$ and $D^-_1(r)$ survive in eqs.(9). 
Surprisingly, but four first order equations for the functions 
$B^+_1(r)$, $D^+_1(r)$, $C^+_1(r)$ and $D^-_1(r)$ reduce to {\it 
one} second order equation. Namely, if
 
 $$ 
 x_+=D^+_1+\sqrt{2} B^+_1,\qquad x_-=D^+_1-\sqrt{2} B^+_1,
 $$
 $$
y_+=D^-_1+\sqrt{2} C^+_1,\qquad y_-=D^-_1-\sqrt{2} C^+_1 
 $$ 
 then simple 
analysis of eqs.(9) shows that either 
 $$ 
 \begin{array}{l} 
 y_- \equiv  0\\
  {y_+}^{''}+({\cal E}^2 -{{2 H^2} \over {r^2}} -{{\sqrt{2} 
                      H^{'}} \over r}){y_+}=0\\
      x_+ =-{1 \over {2r{\cal E}}}y_+ \\ 
      x_- ={1 \over {\cal E}}({y_+}^\prime - ({1 
\over {2r}}+{{\sqrt{2}H} \over r})y_+) \\ \end{array}\eqno(10) $$ or 
 $$
 \begin{array}{l}
 y_+ \equiv 0\\
 {y_-}^{''}+({\cal E}^2 - {{2 H^2} \over {r^2}}
      +{{\sqrt{2} H^{'}} \over r}){y_-}=0\\
 x_-=-{1 \over {2r{\cal E}}}y_- \\ 
 x_+={1 \over {\cal E}}({y_-}^\prime -
({1 \over {2r}}-{{\sqrt{2}H} \over r})y_-) 
 \\
  \end{array}\eqno(11) 
 $$

Eqs.(10) as well as eqs.(11) has  only {\it one} square 
 locally integrable solution at $r=R$. It automatically satisfies 
 boundary condition (12) 
 $$ 
 y_{\pm}(R)=0\eqno(12) 
 $$

 This means that quark cannot penetrate through the sphere $r=R$. 
 Indeed, equations describing quark penetration through  the 
 potential barrier at $r=R$ must have at least two linear 
 independent locally square integrable at $r=R$ solutions that 
 correspond to two possible directions of quark motion (into the 
 sphere and out of the sphere).Lowest eigenvalues 
 ${\cal E}={\cal E}_n/R$ of eqs.(10),(11) with function $H$ 
 defined by eqs.(2),(4) are given in table 1.    

\begin{table}[tbh]
\begin{center}
 \begin{tabular}{|c||c|}
 \hline
 \multicolumn{2}{|c|}{${\cal E}_n$}\\ 
 \hline
 \hline
$   7.800 $ & $  1.997 $ \\
 \hline
$  10.920 $ & $  7.288 $ \\
 \hline
$  14.074 $ & $  10.885 $ \\
 \hline
 \end{tabular}               
\end{center}
\caption{Eigenvalues ${\cal E}={\cal E}_n/R$ for eq.(10) and (11).
 Data in the left and right columns correspond to eq.(10) and (11) 
respectively.}
 \end{table} 

Energy levels can be can be express via numbers ${\cal E}_n$ 
as 
$$
 E_n(R)=\sqrt{ {{\cal E}^2_n \over R^2}+m^2 }\eqno(13)
$$

Let us consider the case $J\ge 3/2$. In this case the equations (9) 
have four linearly independent solutions that are locally square 
integrable at $r=R$. Two of them vanish at $r=R$ whereas two other 
have asymptotics

$$
  \left(
             \begin{array}{c}
                   B^+_1\\B^-_1\\D^+_1\\D^-_1\\C^-_1\\C^+_1
             \end{array}
           \right)
{\longrightarrow \atop r\to R}\qquad
C_1\left( \qquad \left(
                 \begin{array}{c}
                       J-1/2\\
                       0\\0\\0\\
                       J+3/2\\0
                 \end{array}
               \right)
             +
           O(1-r/R)\right)\eqno(14)
$$  
$$
             \left(
             \begin{array}{c}
                   B^+_1\\B^-_1\\D^+_1\\D^-_1\\C^-_1\\C^+_1
             \end{array}
           \right)
{\longrightarrow \atop r\to R}\qquad
 C_2\left( \qquad \left(
                 \begin{array}{c}
                       0\\J+3/2\\
                       0\\0\\0\\
                       J-1/2
                 \end{array}
               \right)
             +
           O(1-r/R)\right)\eqno(15)       
$$  
where $C_1,C_2$ are constants.

This means that quarks with $J\ge 3/2$ are not confined inside the 
sphere $r=R$. In particular, normal component of the carrent 
$j_{\nu}={\bar \psi}\gamma_{\nu}\psi$ on the surface $r=R$ is not 
vanished:
$$
\left. n_{\nu}j^{\nu}\right|_{r=R}\sim C_1C_2\eqno(16)
$$

In such situation the only possibility to confine quark with 
$J\ge 3/2$  sphere $r=R$ is to impose ("by hands",as in usual bag 
models) the boundary condition
$$
\left. n_{\nu}j^{\nu}\right|_{r=R}=0\eqno(17)
$$

Obviously eq.(17)  implies $C_1C_2=0$ and so either $C_1=0$ or 
$C_2=0$. In other words, we must forbid either asymptotics (14) or 
asymptotics (15). We choose the condition $C_1=0$.


Results of numerical evolution of lovest eigenvalues ${\cal E}$ of 
the system (9) with boundary condition $C_1=0$ are presented in 
table 2.

\begin{table}[tbh]
\begin{center}
 \begin{tabular}{|c||c|}
 \hline
 \multicolumn{2}{|c|}{${\cal E}_n$}\\ 
 \hline
 $J=3/2$ & $J=5/2$ \\ 
 \hline
 \hline
$  4.165 $ & $  5.151 $ \\
 \hline
$  6.977 $ & $  8.518 $ \\
 \hline
$  7.949 $ & $  8.863 $ \\
 \hline
 \end{tabular}               
\end{center}
\caption{Eigenvalues ${\cal E}={\cal E}_n/R$ for eq.(9), .
 Data in the left and right columns correspond to $J=3/2$ and 
$J=5/2$ respectively.} 
 \end{table}

\section{Hadron mass spectrum.}

Let  us consider "partition function"
$$
Z=tr\, e^{-iHT}
$$
where $H$ is QCD Hamiltonian. "Partition function" can be 
represented as
$$
Z=\int\, DA\, D{\bar \Psi} D\Psi 
e^{i\int^T_0 dt\int d^3x(L_{YM}(A)+L_{ferm}
({\bar \Psi},\Psi,A))}\eqno(18) 
$$

We assume that the main contribution in functional integral (9) is 
given by trajectories close  to classical solution $A_{cl}$ of YM 
equations defined by eqs.(2),(4),and (6). Applying stationary 
phase method, one gets in zero approximation: 
$$
Z=\int\limits_{zero \atop modes} 
e^{-iE_{YM}(R)T}\prod_q det[\gamma^0 (i\gamma^0
{\partial \over \partial t}+i{\vec \gamma}{\vec \nabla}(A_{cl})-m_q)
]_{APBC} \eqno(19) 
$$
 where $APBC$ means 
"anti-pereodic boundary conditions" and $E_{YM}(R)$ is classical 
energy of the field $A_{cl}$. The determinant in (19) can be easy 
evaluated in terms of positive eigenvalues $E_s(R,m_q)$  
(defined by eq.(13)) of Hamiltonian
$$
i\gamma^0{\vec \gamma}{\vec \nabla}(A_{cl})-\gamma^0m_q
$$
that have been evaluated in section 3 (see, for 
instance,\cite{padj}). Substituting the result in (19), one obtains:  
$$
Z=\sum_{1\le k_s\le n_s}           
\int\limits_{zero \atop modes} 
e^{-i(E_{YM}(R)+\sum_q \sum_s k_sE_s(R,m_q))T} \eqno(20) 
$$
where $n_s=4(2J+1)$ is doubled multiplicity of the eigenvalue $E_s$ 
(Remind, that multipl6icity of the eigenvalue $E_s$  with given $J$ 
is $2(2J+1)$. Additional factor 2 correspond to contribution of 
anti-quarks.).

By virtue of scale invariance of YM equations the measure of 
integration in (20) comprises integration with respect to $R$. So, 
applying again stationary phase method, one gets 
$$
 Z\sim  
\sum_{1\le k_s\le n_s} e^{-i(E_{YM}(R_0)+\sum_q 
\sum_s k_sE_s(R_0,m_q))T} \eqno(22) 
$$
 where $R_0$ is defined from 
 an equation 
$$
 \left.  {\partial \over \partial R} 
[E_{YM}(R)+k_sE_s(R,m_q)]\right|_{R=R_0}=0\eqno(23)
$$
    
So hadronic masses are defined by the formule
$$
M^{theor}=E_{YM}(R_0)+\sum_{quark}E(R_0,m_q)\eqno(24)
$$   
 
The quantity $E_{YM}(R)$ is divergent due to singularity of $A_{cl}$ 
at $r=R$ . We assume that $E_{YM}(R)$ become finite after 
renormalization. However, so far we have not elaborated 
renormalization procedure in our model. Instead, we simply postulate 
that 
$$
E_{YM}(R)=BR^n\eqno(25)
$$ 
where B is some constant. The choice $n=3$ correspond to MIT model. 
In our model the choice $n=2$ also seems natural because the main 
contribution in $E_{YM}(R)$ is proportional to the area of this 
sphere.

Fortunately, it appears that hadronic masses depend on $n$ very 
weakly. So concrete choice of this parameter is not important.

Results of calculation of hadronic masses, corresponding to ground 
state configurations of quarks, are presented in Tables 3 and 4. 
Parameters $B$,$m_u=m_d$,$m_s$, $m_c$ and $m_b$ are determined by 
minimizing of the quantity 
$$ 
\Delta(B,m_u,m_s,m_c,m_b)=  
\sum_{h} 
[(M^{exp}_h-M^{theor}_h)/M^{exp}_h]^2 
$$
 where $M^{exp}_h$ are masses of 
hadrons \cite{pdg}, $M^{theor}_h$ are defined by eq.(24).

To be exact, for determination of the constant B and quark masses we 
used masses of 
$p(n)$, 
 $\Xi$, $\Lambda_c$, $\Lambda^0_b$, $D^{\pm}_c$, 
$D^{\pm}_{sc}$, $B$ and  $\eta_c$ that are measured with the 
best accuracy in comparison with ones of other hadrons.

\begin{table}[thb]
\begin{center}
 \begin{tabular}{|c||c|c|}
 \hline
 Parameters & $ n=3 $ & $ n=2 $ \\ 
 \hline
 \hline
$ B $   & $ 3.338\times 10^{8}(MeV)^4 $ & $ 
2.960\times 10^{6}(MeV)^3 $ \\ \hline $ m_u $ & $ 89.92(MeV) $ & $ 
95.56(MeV) $ \\ \hline $ m_s $ & $ 359.1(MeV) $ & $ 350.4(MeV) $ \\ 
\hline $ m_c $ & $ 1432(MeV) $ & $ 1409(MeV) $ \\ \hline $ m_b $ & $ 
4893(MeV) $ & $ 4866(MeV) $ \\ \hline \end{tabular} \end{center} 
\caption{Parameters of our model.} \end{table}

\begin{table}[h]
\begin{center}
 \begin{tabular}{|c|c||c|c|}
 \hline
Particle(spin,mass(MeV)) & Quark  & Mass $(n=3)$ & Mass 
$(n=2)$ \\ 
& configuration & (MeV) & (MeV) \\
\hline 
\hline
$p(n)\, (J=1/2,m=940)$&$1S_u-1S_u-1S_d$&$957$&$957$\\
 \hline
$\Lambda\, (J=1/2,m=1116)$&$1S_u-1S_u-1S_s$&$1137$&$1137$\\
$\Sigma\, (J=1/2,m=1.189)$&                 &       &       \\
 \hline
$\Xi\, (J=1/2,m=1320)$&$1S_u-1S_s-1S_s$&$1311$&$1311$\\
 \hline
$\Lambda_c\, (J=1/2,m=2285)$&$1S_u-1S_u-1S_c$&$2156$&$2149$\\
$\Sigma_c\, (J=1/2,m=2455)$&                 &       &       \\
 \hline
$\Lambda_b\, (J=1/2,m=5641)$&$1S_u-1S_u-1S_b$&$5601$&$5593$\\
 \hline
$\Xi_c\, (J=1/2,m=2465)$&$1S_u-1S_s-1S_c$&$2326$&$2318$\\
 \hline
$D^{\pm}_c\, (J=0,m=1869)$&$1S_u-1S_c$&$1874$&$1883$\\
 \hline
$D_s\, (J=0,m=1968)$&$1S_s-1S_c$&$2029$&$2033$\\
 \hline
$B\, (J=0,m=5278)$&$1S_u-1S_b$&$5313$&$5320$\\
 \hline
$\eta_c\, (J=0,m=2979)$&$1S_c-1S_c$&$3011$&$3000$\\
 \hline
$\rho\, (J=1,m=769)$&$1S_u-1S_u$&$701$&$721$\\
 \hline
$K^*\, (J=1,m=892)$&$1S_u-1S_s$&$869$&$887$\\
 \hline
$\phi\, (J=1,m=1019)$&$1S_s-1S_s$&$1032$&$1047$\\
 \hline
$\Upsilon\, (J=1,m=9460)$&$1S_b-1S_b$&$9856$&$9830$\\
 \hline
$\Delta^{++}\, (J=3/2,m=1232)$&$1S_u-1S_u-1P_u$&$1191$&$1159$\\
 \hline
$\Omega^{-}\, (J=3/2,m=1672)$&$1S_s-1S_s-1P_s$&$1671$&$1646$\\
 \hline
$\Omega^{0}_c\,
(J=1/2,m=2700(?))$&$1S_s-1S_s-1S_c$&$2491$&$2480$\\
 \hline 
 \end{tabular} 
 \end{center}
\caption{Hadron mass spectrum evaluated by formulas (23)-(25). 
$1S_u-1S_u-1S_s$ means configuration in which two $u$ quarks and one 
$s$ quark in ground states with $J=1/2$, etc; $1P_u$ and $1P_s$ 
means ground states of $u$ and $s$ quark with $J=3/2$. One notes 
that configurations $1S_u-1S_u-1S_u$ and $1S_s-1S_s-1S_s$ are 
forbidden by Pauli principle. So just configurations 
$1S_u-1S_u-1P_u$ and $1S_s-1S_s-1P_s$ are true ground state 
configurations for $\Delta^{++}$ and $\Omega^{-}$. The value of spin 
in these cases automatically equal to 3/2.} \end{table}

\section{Discussion.} We see that our model gives rather good 
description for almost all hadron masses corresponding to ground 
states configurations of quarks except those of light pseudoscalar 
mesons. As we already have mentioned in Introduction, the 
description of the light pseudoscalar mesons cannot be given without 
consideration of the problem of chiral symmetry breaking that is out 
of the scope of our model nowadays.

The accuracy 3-7 percents achieved in our model is maximal possible 
one for any constituent quark model in which interaction between 
quarks is not taken into account. Indeed, in any such model 
$\Lambda$ and $\Sigma$ particle must have the same mass. But really 
there exists approximately 7\% difference between 
$m_{\Lambda}$ and $m_{\Sigma}$.The same is true for 
${\Lambda}_c$ and ${\Sigma}_c$. So seven percents is, most likely, 
the maximal accuracy that can be achieved in any simple constituent 
quark model. In fact, this means that we cannot describe spectrum of 
hadron resonances in framework of our model now. Indeed, mass 
differences between hadron resonances are less, typically, then 
seven percents.

Our model meets some internal difficulties. In particular, we are 
obliged to impose "by hands" boundary condition (8) for quarks with 
$J\ge 3/2$. May be this difficulties (as many others) can be 
overcome if one consider instead of the solution (6) more general 
solution of SU(3) YM equation. Indeed, in the present paper we 
investigate, in fact, SU(2) QCD with quarks in adjoint 
representation. But qauge group of QCD is definitely SU(3) rather 
than SU(2). Most likely, many difficulties arising in our model  
are connected with this circumstance. 

In the nearest future we  plan to investigate singular solutions of 
SU(3) YM equations that cannot be reduced to any solution of SU(2) 
ones and also to develop perturbative theory on the background of 
such singular solutions. Results of the present paper show that 
there exists a good chance to obtain satisfactory description of 
mass spectrum and other properties of hadron in this way.


\begin{thebibliography}{99}
 \bibitem{Bogol}
 P.N. Bogolubov,  Ann.Inst.Henri Poincare 8 (1967) 163

 \bibitem{MIT}
 A. Chodos,R.L. Jaffe,K. Johnson,C.B. Thorn., and V.F. Weisskopf, 
 Phys.Rev. D9 (1974) 3471;
 Phys.Rev. D10 (1974) 2599

 \bibitem{Moddif}
   W.A. Bardeen, M.S. Chanowitz, S.D. Drell,
   M. Weinstein, and T.-M. Yan, Phys.Rev. D11 (1978) 1094;
   P. Hasenfratz, and J. Koti, Phys.Rep. 40C (1978) 75;
   P.N. Bogolubov and A.E. Dorokhov, Fiz. Elem. Chastits At. Yadra
   18 (1987) 917; A.E.Dorokhov, Yu.A. Zubov, and N.I. Kochelev,
   Sov.J.Part.Nucl. 23 (1992) 522 

 \bibitem{Propot}
  T. DeGrand, R.L.Jaffe, K. Johnson, and J. Kiskis,
   Phys.Rev. D12 (1975) 2060;
   K.~Johnson, Acta Phys. Pol. B6 (1975) 865

  \bibitem{Swank} J.M. Swank,L.J. Swank, and T. Dereli, 
                  Phys.Rev., D12 (1975) 1096;
 A.P. Protogenov,  Phys. Lett. B 67 (1977) 62;
                  Phys. Lett. B 87 (1979) 80

  \bibitem{Lunev93} 
 F.A. Lunev,  At. Nucl. Phys. (Yad.Fiz.), 56 (1993) 1591;
                          Phys. Lett. B, 311 (1993) 273;
                          Teor.Math.Phys. 94 (1993) 66

  \bibitem{Lunev92}
  F.A. Lunev,  Phys.  Lett. B 295 (1992) 99.;
  K. Johnson, in {\it QCD-20 years later.}, p.795 (World Scientific, 
   1993);
  M.Bauer, D.Z. Freedman, and P.E. Haagensen, Nucl.Phys. B428 (1994) 147

  \bibitem{Lunev95}
   F.A. Lunev, hep-th/9503133  

  \bibitem{Sing} D. Singlton,  Phys.Rev. D51 (1995) 5911   
   
  \bibitem{Obukhov}
   S. Mahajan and P. Valanju,  
   Phys.Rev., D36 (1987) 1500;     
   Phys.Rev., D36 (1987) 2543;
   Yu.N. Obukhov, hep-th/9608011   
   
 
  \bibitem{wy} T.T. Wu and C.N.Yang, in {\it Properties of mather
  under unusual condition.} eds. H.Mark,S.Fernback 
(Interscience,N.Y.),1969
        
  \bibitem{coo} F. Cooper, A. Khare, and U. Suktme,  
   Phys.Rept., 251 (1995) 267
 

 \bibitem{padj} R. Rajaraman, {\it Solitons and  Instantons.}
   (Noth-Holland Publishing Company, Amsterdam, New York, Oxford, 
   1982)
    
 \bibitem{pdg} Partical data gruop. Phys.Rev. D50 (1994) N3(1)                            
 
  
  
 

 \end{thebibliography}
\end{document}